# *p*-wave triggered superconductivity in single layer graphene on an electron-doped oxide superconductor

A. Di Bernardo[1], O. Millo[2]*, M. Barbone[3], H. Alpern[2], Y. Kalcheim[2], U. Sassi[3], A. K. Ott[3], D. De Fazio[3], D. Yoon[3], M. Amado[1], A. C. Ferrari[3], J. Linder[4], J. W. A. Robinson[1]★

**Electron pairing in the vast majority of superconductors follows the Bardeen-Cooper-Schrieffer theory of superconductivity, which describes the condensation of electrons into pairs with antiparallel spins in a singlet state with an *s*-wave symmetry. Unconventional superconductivity is predicted in single layer graphene where the electrons pair with a *p*-wave or chiral *d*-wave symmetry, depending on the position of the Fermi energy with respect to the Dirac point. By placing single layer graphene on an electron-doped (non-chiral) *d*-wave superconductor and performing local scanning tunnelling microscopy and spectroscopy, here we show evidence for a *p*-wave triggered superconducting density of states in single layer graphene. The realization of unconventional superconductivity in single layer graphene offers an exciting new route for the development of *p*-wave superconductivity using two-dimensional materials with transition temperatures above 4.2 K.**

1. Department of Materials Science and Metallurgy, University of Cambridge, 27 Charles Babbage Road, Cambridge CB3 0FS, United Kingdom.

2. Racah Institute of Physics and the Hebrew University Center for Nanoscience and Nanotechnology, The Hebrew University of Jerusalem, Jerusalem 91904, Israel.

3. Cambridge Graphene Centre, University of Cambridge, Cambridge CB3 0FA, United Kingdom.

4. Department of Physics, Norwegian University of Science and Technology, N-7491 Trondheim, Norway.

Address for correspondence:    ★jjr33@cam.ac.uk
                               * milode@mail.huji.ac.il

# INTRODUCTION

Electron pairing in the vast majority of superconductors follows the Bardeen, Cooper and Schrieffer (BCS) theory of superconductivity[1] which describes the condensation of electron pairs with opposite spins in a singlet state with an isotropic order parameter (*s*-wave symmetry). Superconductivity in a number of compounds is predicted to be unconventional in that the electrons pair in a triplet state with parallel spins and having anisotropic *p*-wave symmetry, most notably in UPt$_3$ (Ref. 2) and Sr$_2$RuO$_4$ (SRO)[3,4]. Although the investigation of compounds that are thought to be bulk *p*-wave superconductors has been intensive over the past two decades (Refs. 5,6) there is still much debate over the nature of the superconductivity in these compounds. Theoretically, a *p*-wave superconducting state is highly sensitive to electron scattering which has meant that, with exception to Ref. 7, all the experiments to date on potential *p*-wave superconductors have focused on single crystal samples. A further complication relates to the typically low transition temperature ($T_c$) of these superconductors (SRO has a $T_c$ below 1.5 K)[5], which, in conjunction with the requirement of extreme purity, prohibits the fabrication of *p*-wave devices. More recently, tunnelling studies[8,9] have suggested that *p*-wave superconductivity can emerge on the surface of topological insulators coupled to *s*-wave superconductors, but the manipulation of these states for device applications is still an open issue. Triplet (*s*-wave) states have also been reported in spectroscopic experiments involving magnetically inhomogeneous *s*-wave superconductor/ferromagnet hybrids[10,11,12].

Electrons in single layer graphene (SLG) are predicted to condense to a superconducting state, either intrinsically by doping[13,14,15,16,17,18,19,20] or by placing SLG on a superconductor with a BCS or a non-BCS pairing symmetry[21,22]. The resulting symmetry depends on the position of the Fermi energy ($E_F$) with respect to the Dirac point. In particular, for $E_F$ shifts of up to 1 eV, *p*-wave[15,16] is predicted. As the doping approaches the van Hove singularity ($E_F$ ~3eV; Ref. 19), a singlet chiral *d*-wave and triplet *f*-wave symmetry are also predicted[17,20], whereas Ref. 18 found dominant chiral *d*-wave superconductivity near van Hove doping and argued that weak coupling superconductivity for doping levels between half-filling and the van Hove density is of Kohn-Luttinger type and likely to be *f*-wave pairing for disconnected Fermi pockets. Ref. 16 predicted that a non-chiral *p*-wave symmetry is favoured for small nearest-neighbour repulsion V (less than 1.1 eV), small onsite interaction U (~8.4 eV) or large doping (above 10%), whereas the chiral *p*-wave state occurs as U or V are increased or the doping level diminishes with respect to the aforementioned values (in pure SLG at half filling U is ~9.3 eV and V ~5.5 eV; Ref. 16). At low density (~ 20%) and including next-nearest neighbour hopping, a chiral *p*-wave state can emerge (Ref. 15). Moreover, the possibility of spin-triplet *s*-wave pairing has been considered in bilayer graphene (Ref. 23).

Although intrinsic superconductivity in SLG has not been observed[24], superconductivity has been induced by doping SLG with Li adatoms[25], intercalating SLG sheets with Ca[26], or by placing SLG on a superconductor[27]. In the latter case, the intrinsic pairing potential for *p*- or chiral *d*-wave superconductivity can be enhanced theoretically[21,22] to the point that a full transition to a superconducting state is triggered and manifested in the SLG superconducting density of states (DoS). Achieving *p*-wave or chiral *d*-wave superconductivity in SLG via a proximity effect would enable the fabrication of devices and allow the full investigation of how these symmetry states can be utilized in cryogenic technology related to spin transport with extremely low dissipation. Additionally, it would be attractive for applications to achieve *p*-wave superconductivity in SLG above 4.2 K.

The standard technique to probe electron pair symmetry involves measuring the voltage ($V$) bias dependence of the non-linear differential conductance ($dI/dV$-$V$) near the superconducting gap edge, which is proportional to the quasiparticle (DoS)[28]. For a conventional superconductor with *s*-wave (zero spin) pairing, $dI/dV$ diminishes below the gap edge in the tunnelling limit. In contrast, unconventional *p*- and chiral *d*-wave symmetries result in zero-energy quasiparticle excitations and surface bound states[29], which manifest themselves in the tunnelling spectra as subgap structures in the DoS, including V-shaped gaps or strong zero-bias conductance peaks (ZBCPs)[30,31,32,33].

The authors of Ref. 27 locally probed the superconducting DoS in SLG on the *s*-wave superconductor Re by scanning tunnelling microscopy (STM). They inferred induced superconductivity in SLG from the observation of a gapped DoS that matched the underlying layer of Re (*s*-wave). The absence of a subgap structure and, therefore, unconventional superconductivity, may indicate a modification of the SLG band structure[34,35] due to the high carrier density of Re ($n_e$ ~4.5 x $10^{23}$/cm$^3$) resulting in significant charge transfer.

An alternative approach to Ref. 27 is to place SLG on an oxide cuprate superconductor as the lower carrier concentrations of these materials (~$10^{20}$/cm$^3$; Ref. 36) will reduce unwanted effects due to doping, such as band structure modifications. However, most oxide-based superconductors have a complex DoS due to their anisotropic superconducting order parameters[30,31,32], which will necessarily complicate the analysis of the SLG DoS spectra, unless care is taken.

$Pr_{2-x}Ce_xCuO_4$ (PCCO) shares similarities with other high-temperature oxide superconducting materials, such as $YBa_2Cu_3O_7$ (YBCO). In particular, it has a *d*-wave symmetry, albeit with a much longer coherence length of ~30 nm[37] compared to that of YBCO of ~2 nm[37]. Tunnelling studies on hole-doped (x < 0.13) PCCO[38] reveal anisotropic superconducting DoS consistent with *d*-wave symmetry with enhancements of the conductance at zero energy as observed on hole-doped YBCO[39]. In contrast, for electron-doped PCCO (x > 0.13), which is the case considered in our experiment where x = 0.15 in PCCO, the superconducting DoS is found to be isotropic[37,40] with a gapped DoS for all tunnelling directions. This is due to the electron-doped PCCO being in the dirty limit[40] meaning $\xi_s > l_e$ ($\xi_s$ is the superconductor coherence length and $l_e$ the electron mean free path).

In this paper we report scanning tunnelling spectroscopy (STS) on SLG placed on the electron-doped cuprate superconductor PCCO in order to investigate the possibility of inducing unconventional superconductivity in SLG such as *p*-wave[15,16] via a superconductor proximity effect. Our results reveal dramatic modifications of the superconducting DoS in SLG compared to the underlying PCCO at 4 K: a mixture of V-shaped gaps, extreme ZBCPs, and split ZBCPs are observed depending on the position of the STM tip. A theoretical analysis proves that the subgap structures match *p*-wave superconductivity with the different symmetry components of the *p*-wave state depending on the surface orientation of the underlying PCCO.

## RESULTS

The PCCO (200-nm-thick) is grown by pulsed laser deposition on (001) oriented $SrTiO_3$ (STO) (see Methods) and has a near-bulk superconducting transition of 20.5 K (see Fig. 1a and Supplementary Fig. 1a). High-angle X-ray diffraction performed on control PCCO films grown in the same conditions as those measured in this study by STM (Fig. 1b and Supplementary Fig. 1b) reveals poor *c*-axis texturing with strong peaks from the (110) family of planes present. In particular, the rocking curves of the (006) diffraction peak of the PCCO films show a spread in the full width at half maximum (FWHM) in the range 0.42°-0.85° (Figs. 1c-1f), which are large values for perfect (001) texturing. These values are comparable to the FWHMs of the diffraction peaks of the (110) family of planes (with differences typically of ~0.3° for neighbouring peaks belonging to the two families of planes) as shown in Supplementary Figs. 1c-1f, which confirms that the PCCO surface is a mixture of (100) and (110) planes. This is in agreement with previous studies[38,40] which have shown that secondary (110) phases would always form during the PCCO growth.

SLG is grown on Cu by chemical vapour deposition, as described in Ref. 42, and transferred onto PCCO/STO, following the procedure in Ref. 43. The analysis of the Raman spectrum acquired on as-grown graphene on Cu (Fig. 1i, grey curve) shows that the 2D peak can be fitted with a single Lorentzian with a full width at half-maximum FWHM(2D) of ~37 $cm^{-1}$, indicating SLG (Ref. 44). The G peak position, Pos(G), and FWHM(G) are 1587 and 20 $cm^{-1}$, respectively, and the intensity ratio, I(2D)/I(G), and area ratio, A(2D)/A(G), are 2.9 and 5.6, respectively. These indicate a doping of ~200 meV for SLG on Cu at room temperature[45,46,47]. The spectrum shows a small D peak with I(D)/I(G)=0.15, corresponding to a defect concentration of $n_D$ ~3.8×$10^{10}$ $cm^{-2}$ (refs. 46 and 48). A compressive biaxial strain of 0.07% is also estimated from the Raman analysis. To compare the SLG quality before and after transfer, the background signal of PCCO on STO is measured under identical conditions (see Fig. 1h and Methods) and subtracted after normalization to the intensity of the Raman peak of STO at ~300 cm$^{-1}$. After transfer, Pos(G) and FWHM(G) are 1581 and 17 cm$^{-1}$, while I(2D)/I(G) and A(2D)/A(G) are 3.4 and 7.1, respectively (Fig. 1i, red curve). This implies a significant reduction in doping to less than 100 meV[45,46,47], with $n_D$ ~2.2x10$^{10}$ cm$^{-2}$ and compressive biaxial strain of 0.08% after subtraction of the PCCO/STO background. Similar values of defect concentration and strain for as-grown SLG on Cu and after its transfer onto PCCO imply homogeneous SLG on PCCO.

To assess doping and sample quality in the same temperature range used for the superconductivity studies, Raman measurements are also performed between 4.2 K and room temperature (Fig. 1j and Supplementary Fig. 2a). The low-temperature Raman data indicate that doping remains below 100 meV and that there are no structural changes compared to room temperature.

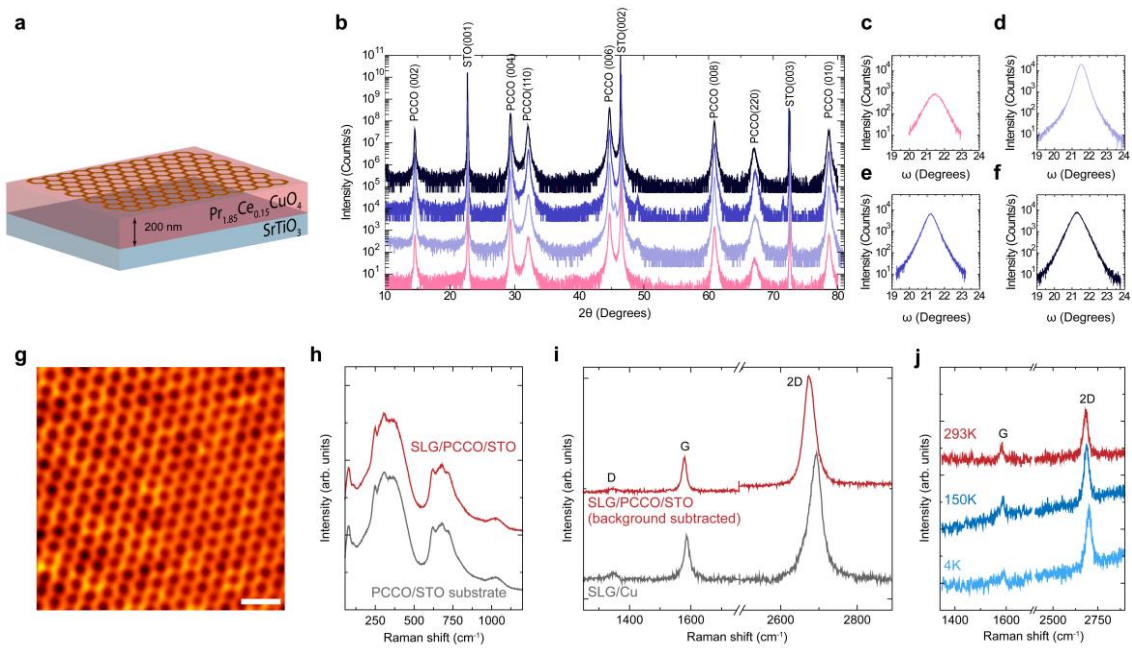

**Figure 1. Characterization of SLG on electron-doped PCCO**. **a**, Schematic of SLG (hexagonal lattice) on PCCO/STO. **b**, High angle X-ray diffraction data on four control films of 5 mm x 5 mm PCCO/STO deposited using identical growth conditions (curves are vertically offset for clarity). **c**, **d**, **e**, **f**, Rocking curves (omega scans) of the (006) PCCO diffraction peaks from the same samples investigated in (**b**) with matching colors showing a full width at half maximum of 0.85° (**c**), 0.42° (**d**), 0.51° (**e**) and 0.61° (**f**). **g**, STM topography map showing SLG on PCCO/STO obtained at 4.2 K (scale bar in **g** has a length of 0.5 nm). **h**, Raman spectra using a laser excitation wavelength of 514.5 nm of PCCO as-grown on STO (grey) and following transfer of SLG (red) at 293 K. **i**, Raman spectra at 514.5 nm of SLG as-grown on Cu (grey) and following transfer onto PCCO/STO (red) measured at 293 K. The PCCO/STO background is subtracted to allow identification of D, G and 2D peaks. **j**, Raman spectra at 514.5 nm of SLG on PCCO/STO, after background subtraction, at 293 K (red), 150 K (blue) and 4.2 K (light blue).

## LOCAL DENSITY OF STATES MEASUREMENTS ON SLG/PCCO

Using STM we locally measure d*I*/d*V*-*V* spectra of SLG on PCCO at 4.2 K and correlate these to surface topography (Fig. 1g). The most predominant superconducting-related spectra show either V-shaped gaps (Fig. 2a) or a subgap structure, including ZBPCs and split ZBCPs (Fig. 2b, c). Additional supporting data are shown in Supplementary Fig. 3. V-shaped gaps were observed in about 45% of the scans, while in all the other areas we observed either ZBCPs (~30%) or split peaks (~25%). We also point out that in the normal state none of these spectral features is observed, which rules out the possibility that these are due to electronic inhomogeneity in the sample, since this should give rise the same features independently on temperature. In some areas, in all of which the SLG local topography cannot be properly resolved, single-electron tunnelling spectra are measured in the superconducting state (Supplementary Fig. 4); such features persisted above $T_c$ (at least up to 50 K), as also shown in Supplementary Fig. 4.

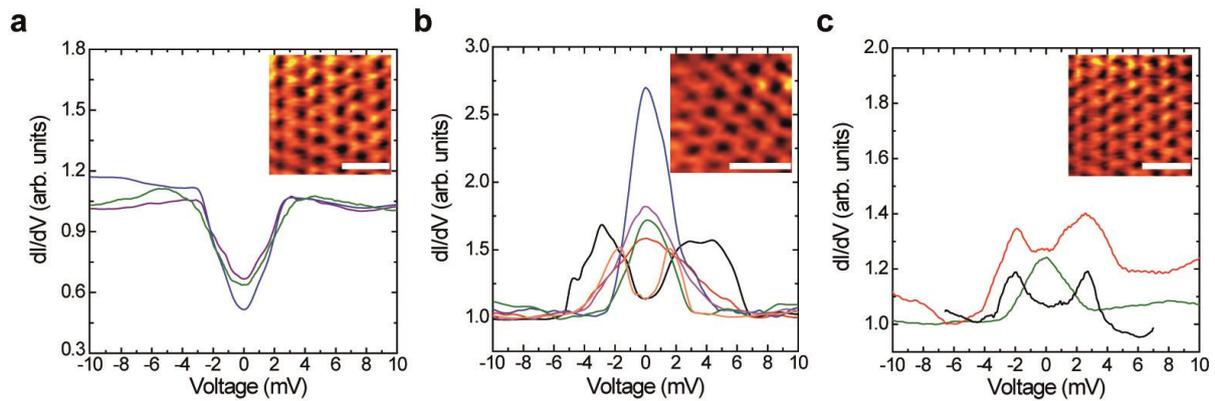

**Figure 2. STM differential conductance versus bias voltage spectra on SLG/PCCO/STO at 4.2 K. a**, **b**, **c**, Typical proximity-induced V-shaped gaps (**a**), ZBCPs (**b**) and split ZBCPs (**c**). Different colors in **a-c** are used to distinguish between spectra recorded in different sample areas. Insets in **a-c** show the typical topography of a sample area where the corresponding spectra in **a-c** are obtained (the scale bars in the insets have a length of 0.5 nm).

The evolution of superconducting–related spectral features is also studied as the sample is warmed up above its superconducting transition. All subgap features including ZBCPs and split ZBCPs are suppressed as the sample is warmed up (Supplementary Fig. 5) giving spectra that are either consistent with lightly doped SLG with $E_F$ within 100 meV from the Dirac point[49] above $T_c$, or to structureless (flat) spectra or single electron tunnelling effects. The latter two appear in less than 10% of the total scanned area, where no superconducting related features are observed below $T_c$ and the STM topographic images do not exhibit a clear SLG structure. This is an important correlation to notice, showing that the proximity takes place only in regions that structurally and electronically conform to the SLG behaviour. Whilst proving that the V-shaped gaps, ZBCPs or split peaks are related to superconductivity, the absence of such structures (particularly conductance peaks) above the superconducting transition temperature, rules out spurious effects from defects (e.g. magnetic impurities) or structural inhomogeneity that may cause Kondo scattering and thus enhancements of the DoS, as such features would also be present in the normal state. Further support for this conclusion comes from the fact that ZBCPs are found (below $T_c$) only in regions where the STM images show clear SLG topography and not in defected regions. We also measure the evolution of ZBCPs in an applied out-of-plane magnetic field (Supplementary Fig. 6), and their magnitude is always found to decrease, with no splitting ever observed, which is also inconsistent with a Kondo effect. This effect of magnetic field is similar to those observed in tunnelling spectroscopy studies of proximity-induced p-wave superconductivity in $Bi_2Se_3$ (Ref. 9) and odd-parity topological superconductivity in $Cu_xBi_2Se_3$ (Ref. 50).

**CONTROL EXPERIMENTS ON BARE PCCO AND Au/PCCO**

Control samples of bare PCCO are also investigated, but no subgap structure or V-shaped gaps are observed (Fig. 3a), even at facets that expose the nodal ab-plane. The only DoS features we can observe are smeared BCS-like gaps, which is consistent with previous STS experiments on electron-doped PCCO[37,40]. We investigate the effect of substrate choice on the superconductor proximity effect by fabricating SLG/PCCO on (001) $LaAlO_3$ (LAO) and observe similar results to STO (Supplementary Fig. 7).

The ZBCPs on SLG/PCCO might be related to the penetration of the anisotropic components of PCCO that are masked in bare PCCO, but which may appear in SLG due to its long electron mean free path (~100 nm near the Dirac point)[51] and spin diffusion length (> 1 μm)[52] being much longer than the coherence length in PCCO of ~30 nm[37]. These are important parameters to consider, since the superconducting condensate is induced in the entire graphene plane. To check this scenario, we replace SLG with Au and fabricate Au/PCCO/STO films (Fig. 3b), where a 10-nm-thick polycrystalline Au layer is deposited in-situ without breaking vacuum by pulsed laser deposition, and perform STS. Au is chosen since it has an electron mean free path ~30 nm at room temperature[53]. The topography maps of Au/PCCO/STO (Fig. 3b) reveal a low surface roughness (~1 nm over a 1 μm$^2$ area; Supplementary Fig. 8), and the corresponding tunnelling spectra on Au show superconducting gaps with no subgap structure, thus supporting our claim that the subgap structures in SLG/PCCO are related to SLG and not the underlying PCCO. We note that the gapped spectra on Au/PCCO are shallower than those of bare PCCO (Fig. 3a, b), but qualitatively similar, suggesting that Au is fully superconducting.

**TUNING THE SUPERCONDUCTING PROXIMITY EFFECT IN SLG/PCCO**

To further strengthen our claim that the spectral features on SLG/PCCO are related to a superconductor proximity effect in SLG and not the underlying PCCO, we also measure spectra on SLG covered with evaporated microislands (10 μm in diameter and 30 nm in height) of either Ag or Au (Fig. 3c, d). Unlike metals, such as Pd, which have a stronger binding interaction with SLG[34], the low binding energy of Au and Ag (0.04 eV per carbon atom[34]) results in a reduced modification of the SLG band structure[34]. Ag (Au) has a work function of 4.4 eV (5.2 eV) which is lower (higher) than the work function of SLG (~4.6 eV; Ref. 34). Therefore, if a superconducting proximity effect occurs on SLG/PCCO, this should be stronger near a Ag microisland than a Au one, since Ag (Au) acts as a donor (acceptor) of electrons in SLG, with a doping profile that can extend up to hundreds nm from the microisland step edge[54]. The tens of d$I$/d$V$-$V$ spectra acquired on these samples show that pronounced ZBCPs are observed near and even on the Ag microislands, but no ZBCPs are ever observed near or on Au microislands, despite the well-defined SLG topography (some representative spectra are shown in Fig. 3c and in Supplementary Figs. 9, 10). This implies that the Au islands reduce the SLG electron density to the extent that the superconducting proximity effect for SLG on PCCO is dramatically suppressed. We also note that these data indicate that possible doping inhomogeneities can only affect the size of the unconventional superconducting features (even to their suppression, along with the proximity effect altogether), but do not affect the unconventional nature of the proximity induced superconductivity in SLG.

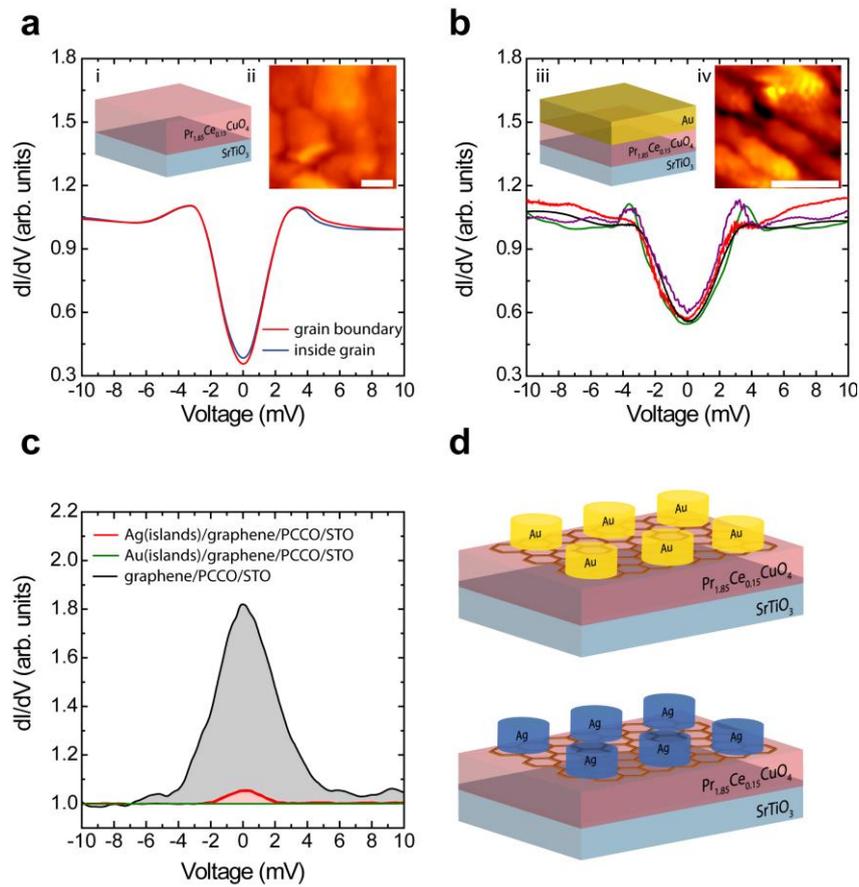

**Figure 3. STM differential conductance versus bias voltage spectra on control samples of PCCO/STO and Au/PCCO/STO at 4.2 K. a**, Typical STM spectra at 4.2 K for PCCO/STO with schematic of the sample structure (inset i) and STM topography (inset ii; the scale bar has a length of 0.5 nm). **b**, Typical STM spectra at 4.2 K for Au/PCCO/STO (Au is 10-nm-thick) with schematic of the sample structure (inset iii) and STM topography (inset iv; the scale bar has a length of 0.5 nm). **c**, **d**, Average ZBCP amplitudes for SLG/PCCO/STO (black shaded curve) and Au (green shaded curve) and Ag (red shaded curve) microislands on SLG/PCCO/STO (**c**) with corresponding schematics shown in (**d**) where the hexagonal lattice represents SLG.

**CALCULATION OF THE SUPERCONDUCTING DOS IN SLG/PCCO**

We now consider the spatial variations in the superconducting DoS that we observe on SLG/PCCO (Fig. 2 and Supplementary Fig. 3). These indicate local variations in the proximity effect due to a combination of PCCO faceting and changes in the PCCO surface that is a mixture of different PCCO planes. In Fig. 4 we have plotted the theoretical DoS on SLG as a function of the STM orientation relative the graphene surface, which varies due to the PCCO faceting.

These are calculated by applying the model in Refs. 22,55, which predicts an effective *p*-wave pairing at the Dirac points in SLG on a *d*-wave superconductor and involves solving a tight-binding Hamiltonian for SLG assuming that the tunnelling between SLG and the STM tip obeys an extended Blonder-Tinkham-Klapwijk (BTK) theory[30,56]. Following Refs. 22,55, one can write the Hamiltonian as follows in a band-basis where $c_q$ is a Fermion-operator in the conduction band and $d_q$ is a Fermion-operator in the valence band:

$$H = \sum_{q,\sigma}[(t\epsilon_q - \mu)c^+_{q\sigma}c_{q\sigma} + (-t\epsilon_q - \mu)d^+_{q\sigma}d_{q\sigma}] - \sum_{q,\sigma}[\delta_q(d^+_{q\uparrow}d^+_{-q\downarrow} - c^+_{q\uparrow}c^+_{-q\downarrow}) + i\, u_q(c^+_{q\uparrow}d^+_{-q\downarrow} - d^+_{q\uparrow}c^+_{-q\downarrow}) + \text{h.c.}] \quad (1)$$

In equation (1) we define the normal-state band dispersion as $\epsilon_q = \sum_a e^{iqa}$ where the sum is over nearest-neighbor vectors *a*, *t* is the hopping parameter, *μ* is the chemical potential, while $\delta_q$ and $u_q$ are associated to the superconducting order parameter. The above Hamiltonian can be diagonalized and yields eigenvalues:

$$E_q = \pm\sqrt{[t\epsilon_q \pm \sqrt{(\mu^2 + u_q^2)}]^2 + \delta_q^2} \quad (2)$$

Since the shift in Fermi level ($E_F$) in SLG (below 100 meV in SLG on PCCO at 4.2 K) is much larger than the superconducting order parameter (Δ ~5meV for PCCO), the quantity $u_q$ in equation (2) can be neglected and we are left with an effective superconductor with gap $\delta_q$. It is this gap that acquires a *p*-wave symmetry near the Dirac points when *d*-wave superconducting correlations are induced in SLG[22,55]. This is due to the fact that the transport properties of SLG are determined by its behaviour near the Dirac points for doping levels comparable to that in our study (< 100 meV) and, as explained in Refs. 22,55, a *d*-wave symmetry in the full Brillouin zone corresponds to a *p*-wave symmetry in the vicinity of the Dirac points at K and K'.

In this regime, where the shift in $E_F$ in SLG is much larger than Δ, a simplified model for tunnelling between SLG and the STM-tip based on BTK theory can be adopted to account for *p*-wave pairing. This is a commonly-used modelling tool for tunnelling measurements with unconventional superconductors[57,58,59]. The procedure consists in setting up the wavefunctions on SLG and on the normal side of the tunnelling barrier, the strength of the latter being characterized by a dimensionless number Z. The Z parameter describes the strength of the tunnel barrier between SLG and the STM tip, i.e. the barrier is effectively the vacuum in between. The Z value affects the DoS spectra as it is related to the ratio between Andreev reflections and quasiparticle tunnelling (Z = 0 corresponds to a perfect interface with zero resistance, whereas Z >> 1 corresponds to a barrier with high tunnel resistance). The Z value can be controlled in STS via the current and bias voltage settings (see below), which governs the tip-sample distance in the case of an ideally homogeneous sample. However, in our experiment, it is realistic to expect spatial variations in Z (for a given current-bias setting) since the STM tip approaches the graphene surface at slightly different distances at different locations, due to variations in the surface plane of the underlying PCCO and local variations in graphene-PCCO connectivity, which is hence accounted in our theoretical model.

The theoretical DoS in Fig. 4 were obtained with $\Delta$ = 5 meV, $\delta_q$ = 0.1 meV and Z varying between 0.8 and 1.7, and they show that the surface orientation of PCCO sensitively controls the DoS, due to the projection of the $p_x$ and $p_y$ symmetry components of the $p$-wave state. Importantly, in this Z regime, enhanced conductance at zero bias, and therefore the ZBCPs, cannot be related to Andreev reflections unless the order parameter is unconventional, meaning it does not have an $s$-wave symmetry. In addition, the sharp structure of the peaks that we observe experimentally (Fig. 2 and Supplementary Fig. 3) can be accounted for only by sign-changing order-parameter symmetry, such as $p$-wave.

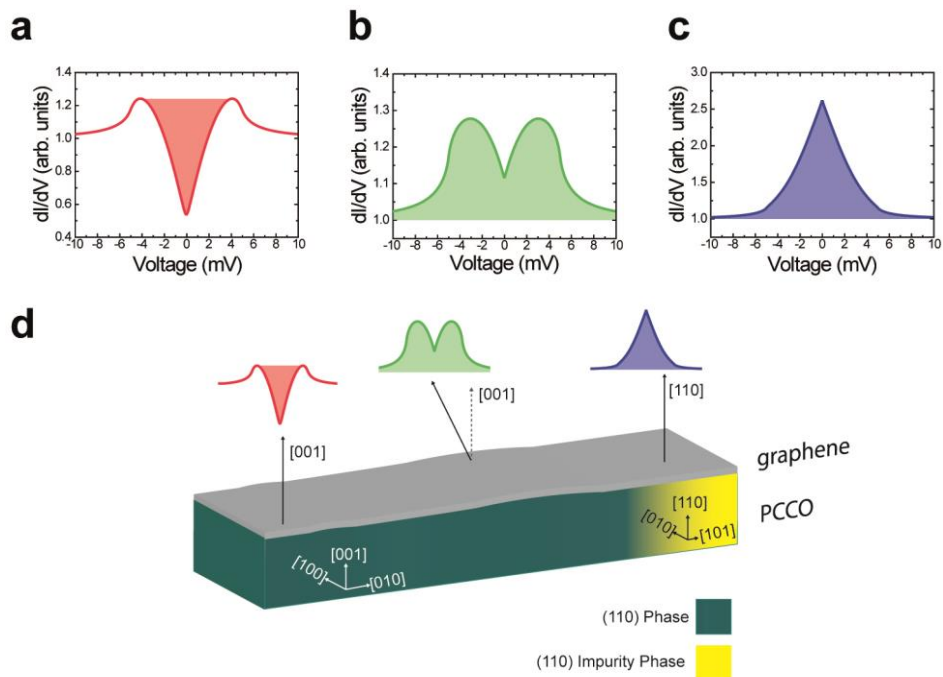

**Figure 4. Theoretical proximity-induced superconducting DoS in SLG on PCCO. a**, **b**, **c**, **d**, The plots show that the projections of the $p$-wave induced symmetries in SLG are dependent on the crystal orientation of the underlying PCCO. This is clearest in (**d**) which illustrates the relationship between the observed DoS on SLG and the plane normal to the underlying PCCO: on flat regions oriented along the [001] crystallographic direction, the DoS corresponds to $p_y$-wave or antinodal $d_{x^2-y^2}$-wave with a strong tunnelling barrier (red curve in **a**); rough areas expose a mixture of normal planes oriented between the [001] and [110] crystallographic directions, where the DoS corresponds to $p_y$-wave with a moderate tunnelling barrier (green curve in **b**); the DoS corresponding to the $p_x$-wave is projected on the phase oriented along the [110] crystallographic direction (blue curve in **c**).

# DISCUSSION

From our results, we conclude that the modifications of the superconducting DoS on SLG/PCCO/STO (Fig. 2 and Supplementary Fig. 3) compared to PCCO/STO (Fig. 3a) or Au/PCCO (Fig. 3b), in particular the ZBCPs and split peaks, indicate the induction of *p*-wave[15,16] or chiral *d*-wave pairing in SLG[20]. However, chiral *d*-wave requires dominant electron-electron repulsion in SLG at much higher doping levels (up to 3 eV; Ref. 20) than here (~100 meV). In our low doping regime, *p*-wave pairing should theoretically dominate[15,16,22] with spectral features in the DoS that match our findings in Fig. 2.

We also exclude other unconventional pairing symmetries that may be compatible with the spectral features we observe such as *f*-wave pairing as suggested in Ref. 60, since this would require a buckling in the honeycomb lattice, which we do not expect for SLG. An interesting possibility, however, would be to consider the emergent order parameter symmetry near the Dirac points in the presence of f-wave pairing[17,18] in the Brillouin zone to determine if this could also provide zero-energy states.

Our theoretical model shows that an induced $p_x$-wave state is manifested in the SLG DoS as a ZBCP (Fig. 4b), while an induced $p_y$-wave state manifests as a split-ZBCP (Fig. 4c). If tunnelling occurs along the *c*-axis of the PCCO (along [001]), a V-shaped gapped DoS consistent with $p_y$ or antinodal $d_{x^2-y^2}$ symmetry is observed (Fig. 4a).

The theoretical analysis is consistent with the distribution in the DoS spectral features that we measure on the SLG/PCCO (~45% of V-shaped gaps, ~30% of ZBCPs and ~25% of split peaks). Although it is not possible to determine the orientation of the underlying PCCO below the STM tip during measurements, the X-ray data in Fig. 1b proves that the surface of PCCO is a mixture of the (001) and (110) family of planes, and therefore the DoS according to our theory should be a mixture of V-shaped gaps (occurring for *c*-axis tunnelling), ZBCPs or split peaks (occurring for tunnelling from the *ab*-plane). We point out that the surface roughness of the PCCO will also expose different crystal orientations. Therefore, the spectroscopic features on SLG/PPCO will be related to a combination of PCCO crystal structure and surface roughness. Fig. 4 also shows that the V-shaped DoS calculated for SLG/PCCO does not reach zero, which is consistent with our experimental observations (Fig. 2a and Supplementary Fig. 3), but also with previous tunnelling experiments on PCCO even at temperatures much lower (1.8 K) than our measurement setup of 4.2 K[40].

Local variations in the Z parameter can determine the amplitude of subgap features (ZBCPs or split peaks) and the depth of V-shaped gaps, as shown in Supplementary Fig. 11. Although Z values can be changed intentionally by varying the set current and set bias-voltage values (before disabling the feedback loop and acquiring a d*I*/d*V* spectrum), care was taken to work with low set currents (~0.1-0.3 nA) and bias voltages just above the gap region (8-10 mV), yielding junction resistances of around 5 x $10^7$ Ω, much larger than the quantum resistance h/$e^2$ (~25.8 kΩ) and thus well within the tunnelling regime. In this regime, the overall spectral features (ZBCPs, split ZBCPs, or V-shaped gaps) measured at a specific location on SLG/PCCO do not change upon varying the current and voltage set values, ruling out the possibility that these features are related to single electron tunnelling effects[61]. It should be noted, however, that the underlying orientation of the PCCO film and the degree of local PCCO-SLG electrical connectivity can also influence the tip-SLG distance and hence the Z value (for a given current-bias setting). We exclude that variations in the spectral features are due to defects in SLG, since STM images do not reveal any structural inhomogeneity in SLG and no damage after transfer (consistent with the Raman analysis), and, in particular, no defects are observed in regions where the spectra are acquired. Further, spectral anomalies disappear in the normal state.

Finally, we point out that in Ref. 62, subgap features in the density of states including ZBCPs and split zero-bias peaks, could be obtained via Andreev bound states formed under a small Pb slab coupled to SLG. Such a system effectively constituted a quantum dot coupled to superconducting leads. In the present work, the local and random doping inhomogeneities in SLG on PCCO are not likely to induce well defined quantum dots, where confinement would be further suppressed by Klein tunnelling, and thus cannot account for the subgap features seen in our STM-data.

Although our analysis is consistent with the triggering of an unconventional $p$-wave order parameter in SLG, we cannot determine the exact symmetry of the $p$-wave state. Since the DoS spectra look qualitatively different from those reported for a chiral spin-triplet $p$-wave state in a topological superconductor[8,9], we believe that our results are consistent with an effective $p$-wave order parameter emerging at the Dirac points in SLG as a result of Andreev bound states induced by the proximity coupling with the $d$-wave pairing potential in PCCO[63]. The results therefore create the exciting possibility of creating $p$-wave superconducting devices on single layer graphene in which quantum coherent electron states, gate tunability, and spin/charge could be addressed in the superconducting state.

Measurements of quasiparticle interference (QPI) modulations[64] could provide further insights into the exact pairing symmetry induced in the SLG on PCCO. Our results suggest that the superconducting pairing of the $p$-wave state in our system is spin-singlet, which means that the Andreev bound-states providing the zero-energy peak seen in the STM data can be interpreted as an odd-frequency spin-singlet odd-parity state[65,66]. Interfacial spin-orbit coupling due to broken inversion symmetry at the PCCO/SLG interface may also in principle induce a triplet component in the system[67]; however, the induction of a triplet component would necessary require significant spin-orbit coupling at the SLG/PCCO interface which is unlikely here due to the low atomic numbers of the elements in PCCO and graphene. Further investigation is required to confirm this.

# Methods

**PCCO film growth.** 200-nm-thick PCCO thin films are grown by pulsed laser deposition on (001) STO using a stoichiometric target fabricated by a solid-state reaction method from high purity $Pr_6O_{11}$ (99.9% purity), $CeO_2$ (99.9% purity) and $CuO$ (99.99% purity) powders. The growth is carried out at 820 °C by firing a Lambda Physik KrF excimer laser ($\lambda = 248$ nm) on a rotating PCCO target with a pulse rate of 7 Hz and energy density of 1.5 J/cm$^2$, after introducing 230 mTorr of $N_2O$ in a ultra-high vacuum chamber (base pressure of $10^{-8}$ mbar). To obtain an optimal superconducting transition (20.5 K; Supplementary Fig. 1a), the PCCO films are annealed in-situ[41] by evacuating the chamber immediately after growth and holding the substrate at 720 °C for 4 minutes. Films are then cooled to room temperature in a vacuum of $10^{-5}$ mbar within 2.5 hours.

**Growth and transfer of SLG on PCCO.** SLG is grown by chemical vapour deposition (CVD) on a 35-μm-thick Cu foil loaded into a hot wall tube furnace, which is subsequently evacuated to a pressure of about 1 mTorr[42]. The Cu foil is annealed in a hydrogen atmosphere ($H_2$, 20 sccm) at 1000 °C for 30 minutes to reduce the Cu oxide component and to increase the grain size[42,68]. SLG growth is then initiated by adding 5 sccm $CH_4$ to the $H_2$ flow. After 30 minutes of growth, the substrate is cooled in vacuum (1 mTorr) to room temperature and then unloaded from the reactor.

CVD is chosen over micromechanical exfoliation[43,69] so as to cover large areas (5x5 mm$^2$) of PCCO with SLG. This is beneficial in our experiments for a number of reasons: the localization and alignment of the sample is simplified (the scan size of our STM is limited to 1x1 μm$^2$, and the optical contrast of SLG on PCCO does not allow for an easy identification of SLG); the large area allows us to fabricate large arrays of metallic islands (covering 2x2 mm$^2$ of the sample surface); the measurements benefit from a larger statistical sample and the presence of local defects can be easily avoided, thus not disturbing tunnelling-current measurements, which only require nm-size areas to be performed.

The CVD-grown SLG is then transferred onto 200-nm-thick epitaxial films of electron-doped PCCO on (001) STO. To do so, a 500-nm-thick layer of polymethyl methacrylate (PMMA) is spin coated onto the SLG/Cu sample. The PMMA/SLG/Cu stack is subsequently immersed in an aqueous solution of ammonium persulfate (APS) to etch the Cu foil[42,43]. The PMMA/SLG stack is then placed in de-ionized (DI) water to rinse off acid residuals and fished off the DI bath using the PCCO substrate. Finally, the PMMA/SLG/PCCO sample is dried at room temperature and placed in acetone to remove the PMMA layer, leaving SLG on PCCO.

**Raman measurements.** Raman spectroscopy is used to investigate the quality and uniformity of the as-grown SLG on Cu as well as to quantify the presence of defects and evaluate doping. Raman spectra are recorded at 514.5 nm using a Renishaw InVia spectrometer equipped with a Leica DM LM microscope and a 100X objective (numerical aperture NA=0.85). Under these conditions the laser spot size is ~1μm. The laser power is kept at ~500 μW to avoid any possible sample damage.

To assess the doping and quality of SLG as a function of temperature, in particular below the superconducting critical temperature of PCCO of 20.5 K, low temperature Raman measurements are performed between room temperature and 4.2 K (Supplementary Fig. 2a). These are carried out using an Oxford Instrument cryostat coupled with a Horiba Jobin Yvon HR Evolution Raman spectrometer. An Olympus LUCPlanFL N 40X objective with an aberration correction ring is used. The aberration correction is adjusted to match the glass window thickness of the cryostat in order to enhance the signal-to-noise ratio.

Each sample is fixed to the holder in the cryostat with vacuum grease (Apiezon N Grease) and the cryostat is pumped to $10^{-6}$ Torr before measurements, in order to reproduce the same vacuum conditions used in the STM. The sample is cooled to 4.2 K using liquid helium and the temperature is raised stepwise with a programmable temperature controller. For each temperature, a Raman spectrum of SLG on PCCO is recorded as well as a reference spectrum on a bare PCCO film on (001) STO. To reveal the SLG contribution to the Raman signal, the spectrum measured on PCCO at a given temperature is subtracted from the corresponding one on SLG/PCCO.

**Data availability**
The dataset generated and analysed during this study are available for access at http://dx.doi.org/10.17863/CAM.6228.

# References

1. Bardeen, J., Cooper, L. N. & Schrieffer, J. R. Theory of superconductivity. *Phys. Rev.* **108**, 1175 (1957).
2. Tou, H. *et al.* Odd-parity superconductivity with parallel spin pairing in UPt$_3$: evidence from $^{195}$Pt Knight shift study. *Phys. Rev. Lett.* **77**, 1374–1377 (1996).
3. Maeno, Y. *et al.* Superconductivity in a layered perovskite without copper. *Nature* **372**, 532-534 (1994).
4. Ishida, K. *et al.* Spin-triplet superconductivity in Sr$_2$RuO$_4$ identified by $^{17}$O Knight shift. *Nature* **396**, 658-660 (1998).
5. Maeno, Y., Kittaka, S., Nomura, T., Yonezawa, S. & Ishida, K. I. Evaluation of spin-triplet superconductivity in Sr$_2$RuO$_4$. *J. Phys. Soc. Jpn.* **81**, 011009 (2012).
6. Linder, J. & Robinson, J.W.A. Superconducting spintronics. *Nat. Phys.* **11**, 307-315 (2015).
7. Krockenberger, Y. *et al.* Growth of superconducting Sr$_2$RuO$_4$ thin films. *Appl. Phys. Lett.* **97**, 082502 (2010).
8. Xu, J. P. *et al.* Experimental Detection of a Majorana Mode in the core of a Magnetic Vortex inside a Topological Insulator-Superconductor Bi$_2$Te$_3$/NbSe$_2$ Heterostructure. *Phys. Rev. Lett.* **114**, 017001 (2015).
9. Koren, G., Kirzhner, T., Kalcheim, Y. & Millo, O. Signature of proximity-induced p$_x$ + ip$_y$ triplet pairing in the doped topological insulator Bi$_2$Se$_3$ by the *s*-wave superconductor NbN. *Europhys. Lett.* **103**, 67010 (2013).
10. Di Bernardo, A. *et al*. Signature of magnetic-dependent gapless odd frequency states at superconductor/ferromagnet interfaces. *Nat. Comm.* **6**, 8053 (2015).
11. Kalcheim, Y., Millo, O., Di Bernardo, A., Pal, A. & Robinson, J.W.A. Inverse proximity effect at superconductor-ferromagnet interfaces: evidence for induced triplet pairing in the superconductor. *Phys. Rev. B* **92**, 060501 (R) (2015).
12. Di Bernardo, A. *et al*. Inverse paramagnetic Meissner effect due to s-wave odd-frequency superconductivity. *Phys. Rev. X* **5**, 041021 (2015).
13. Kopnin, N. B. & Sonin, E. B. BCS superconductivity of Dirac electrons in graphene layers. *Phys. Rev. Lett.* **100**, 246808 (2008).
14. Uchoa, B. & Castro Neto, A. H. Superconducting States of Pure and Doped Graphene. *Phys. Rev. Lett.* **98**, 146801 (2007).
15. Ma, T., Yang, F., Yao, H. & Lin, H.Q. Possible triplet *p+ip* superconductivity in graphene at low filling. *Phys. Rev. B* **90**, 245114 (2014).
16. Faye, J.P.L., Sahebsara, P. & Sénéchal, D. Chiral superconductivity on the graphene lattice. *Phys. Rev. B* **92**, 085121 (2015).
17. Kiesel, M. L., Platt, C., Hanke, W., Abanin, D. A: & Thomale, R. Competing many-body instabilities and unconventional superconductivity in graphene, *Phys. Rev. B* **86**, 020507 (R) (2012).
18. Nandkishore, R., Thomale, R. & Chubukov, A. V. Superconductivity from weak repulsions in hexagonal lattice systems. *Phys. Rev. B* **89**, 144501 (2014).
19. González, J. Kohn-Luttinger superconductivity in graphene. *Phys. Rev. B* **78**, 205431 (2008).
20. Nandkishore, R., Levitov, L. S. & Chubukov, A. V. Chiral superconductivity from repulsive interactions in doped graphene. *Nat. Phys.* **8**, 158-163 (2012).
21. Black-Schaffer, A. M., Doniach, S. Possibility of measuring intrinsic electronic correlations in graphene using a d-wave contact Josephson junction. *Phys. Rev. B* **81**, 014517 (2010).
22. Linder, J., Black-Schaffer, A. M., Yokoyama, T., Doniach, S. & Sudbø, A. Josephson current in graphene: Role of unconventional pairing symmetries. *Phys. Rev. B* **80**, 094522 (2009).
23. Hosseini, M. V. & Zareyan, M. Model of an Exotic Chiral Superconducting Phase in a Graphene Bilayer. *Phys. Rev. Lett.* **108**, 147001 (2012).
24. Kotov, V. N., Uchoa, B., Pereira, V. M., Guinea, F. & Castro Neto, A. H. Electron-electron interactions in graphene: current status and perspectives. *Rev. Mod. Phys.* **84**, 1067-1125 (2012).
25. Ludbrook, B. M. *et al*. Evidence for superconductivity in Li-decorated monolayer graphene. *Proc. Natl. Acad. Sci.* **112**, 11795-11799 (2015).
26. Chapman, J. *et al*. Superconductivity in Ca-doped graphene laminates, *Sci. Rep.* **6**, 23254 (2016).
27. Tonnoir, C. *et al.* Induced superconductivity in graphene grown on Rhenium. *Phys. Rev. Lett.* **111,** 246805 (2013).
28. Wolf, L. E. *Principles of Electron Tunnelling Spectroscopy* (Oxford Univ. Press, 2012).
29. Hu, C. R. Midgap surface states as a novel signature of d$_{xa2-xb2}$ wave superconductivity. *Phys. Rev. Lett.* **72**, 1526-1529 (1994).
30. Tanaka, Y. & Kashiwaya, S. Theory of tunnelling spectroscopy of *d*-wave superconductors. *Phys. Rev. Lett.* **74**, 3451-3454 (1995).
31. Kashiwaya, S. & Tanaka, Y. Tunnelling effects on surface bound states in unconventional superconductors. *Rep. Prog. Phys.* **63**, 1641-1724 (2000).

32. Yamashiro, M. & Tanaka, Y. Theory of tunnelling spectroscopy in superconducting $Sr_2RuO_4$. *Phys. Rev. B* **56**, 7847-7850 (1997).
33. Li, X. W. Tunnelling conductance in normal metal/insulator/triplet superconductor junctions. *Comm. Theo. Phys.* **44**, 381-384 (2005).
34. Khomyakov, P. A. *et al.* First-principles study of the interaction and charge transfer between graphene and metals. *Phys. Rev. B* **79**, 195425 (2009).
35. Wintterlin, J. & Bocquet, M. L. Graphene on metal surfaces. *Surf. Sci.* **603**, 1841-1852 (2009).
36. Li, P., Balakirev, F. F. & Greene, R. L. High-field Hall resistivity and magnetoresistance of electron-doped $Pr_{2-x}Ce_xCuO_{4-\delta}$. *Phys. Rev. Lett.* **99**, 047003 (2007).
37. Kalcheim, Y., Millo, O., Egilmez, M., Robinson, J. W. A. & Blamire, M. G. Evidence for anisotropic triplet superconductor order parameter in half-metallic ferromagnetic $La_{0.7}Ca_{0.3}Mn_3O$ proximity coupled to superconducting $Pr_{1.85}Ce_{0.15}CuO_4$. *Phys.Rev. B* **85**, 104504 (2012).
38. Qazilbash, M. M., Biswas, A., Dagan, Y., Ott, R. A. & Greene, R. L. Point-contact spectroscopy of the electron-doped cuprate superconductor $Pr_{2-x}Ce_xCuO_4$: the dependence of conductance-voltage spectra on cerium doping, barrier strength, and magnetic field. *Phys. Rev. B* **68**, 024502 (2003).
39. Sharoni, A. *et al.* Scanning tunnelling spectroscopy of a-axis $YBa_2Cu_3O_{7-\delta}$ films: k-selectivity and the shape of the superconductor gap. *Europhys. Lett.* **62**, 883-889 (2003).
40. Dagan, Y., Beck, R. & Greene, R. L. Dirty Superconductivity in the Electron-Doped Cuprate $Pr_{2-x}Ce_xCuO_{4-\delta}$: Tunnelling Study. *Phys. Rev. Lett.* **99**, 147004 (2007).
41. Maiser, E. *et al.* Pulsed-laser deposition of $Pr_{2-x}Ce_xCuO_{4-y}$ thin films and the effect of high-temperature post-annealing. *Physica C* **297**, 15-22 (1998).
42. Bae, S. *et al.* Roll-to-roll production of 30-inch graphene films for transparent electrodes. *Nat. Nanotechnol.* **5**, 574-578 (2010).
43. Bonaccorso, F. *et al.* Production and processing of graphene and 2d crystals. *Mater. Today* **15**, 564-589 (2012).
44. Ferrari, A. C. *et al.* Raman spectrum of graphene and graphene layers. *Phys. Rev. Lett.* **97**, 187401 (2006).
45. Das, A. *et al.* Monitoring dopants by Raman scattering in an electrochemically top-gated graphene transistor. *Nat. Nanotechnol.* **3**, 201-215 (2008).
46. Bruna, M. *et al.* Doping dependence of the Raman spectrum of defected graphene. *ACS Nano* **8**, 7432-7441 (2014).
47. Pisana, S. *et al.* Breakdown of the adiabatic Born-Oppenheimer approximation in graphene. *Nat. Mater.* **6**, 198-201 (2007).
48. Cancado, L. G. *et al.* Quantifying defects in graphene via Raman spectroscopy at different excitation energies. *Nano Lett.* **11**, 3190-3196 (2011).
49. Xue, J. *et al.* Scanning tunnelling microscopy and spectroscopy of ultra-flat graphene on hexagonal boron nitride. *Nat. Mat.* **10**, 282-285 (2011).
50. Sasaki, S. *et al.* Topological superconductivity in $Cu_xBi_2Se_3$. *Phys. Rev. Lett.* **107**, 217001.
51. Giannazzo, F., Sonde, S., Lo Nigro, R., Rimini, E. & Ranieri, V. Mapping of the density of scattering centres limiting the electron mean free path in graphene. *Nano Lett.* **11**, 4612-4618 (2011).
52. Yan, W. *et al.* Long-spin diffusion length in few-layer graphene flakes, *Phys. Rev. Lett.* (2016), in press.
53. Jen, S.U., Yu, C.C., Liu, C.H. & Lee, G.Y. Piezoresistance and electrical resistivity of Pd, Au, and Cu films. *Thin Solid Films* **434**, 316 (2003).
54. Giovannetti, G. *et al.* Doping graphene with metal contacts. *Phys. Rev. Lett.* **101**, 026803 (2008).
55. Jiang, Y., Yao, D.-X., Carlson, E. W., Chen, H.-D. & Hu, J. P. Andreev conductance in the d+id –wave superconducting states of graphene. *Phys. Rev. B* **77**, 235420 (2008).
56. Blonder, G. E., Tinkham, M. & Klapwijk, T. Transition from metallic to tunnelling regimes in superconducting microconstrictions: excess current, charge imbalance, and supercurrent conversion. *Phys. Rev. B* **25**, 4515 (1982).
57. Wei, J. Y. T., Yeh, N. C., Garrigus, D. F. & Strasik, M. Directional tunnelling and Andreev reflection on $YBa_2Cu_3O_{7-\delta}$ single crystals: predominance of d-wave pairing symmetry verified with the generalized Blonder, Tinkham and Klapwijk theory. *Phys. Rev. Lett.* **81**, 2542 (1998).
58. Daghero, D. *et al.* Strong-coupling d-wave superconducting in $PuCoGa_5$ probed by point-contact spectroscopy. *Nat. Commun.* **3**, 786 (2012).
59. Aggarwal, L. *et al.* Unconventional superconductivity at mesoscopic point contacts on the 3D Dirac semimetal $Cd_3As_2$. *Nat. Mater.* **15**, 32-37 (2016).
60. Zhang, L. D., Yang, F. & Yao, Y. Possible electric-field-induced superconducting states in doped silicene, *Sci. Rep.* **5**, 8203 (2015).
61. Hanna, A. E. & Tinkham, M. Variation of the Coulomb staircase in a two-junction system by fractional electron charge, *Phys. Rev. B* **44**, 5919 (R) (1991).

62. Dirks, T. *et al.* Transport through Andreev bound states in a graphene quantum dot, *Nat. Phys.* **7**, 386-390 (2011).
63. Tanaka, Y., Sato, M. & Nagaosa, N. Symmetry and topology in superconductors – odd-frequency pairing and edge states. *J. Phys. Soc. Jpn.* **81**, 011013 (2012).
64. M. H. Hamidian *et al.*, Atomic-scale electronic structure of the cuprate *d*-symmetry form factor density wave state, *Nat. Phys.* **12**, 150-156 (2016).
65. Tanaka,Y., Golubov, A., Kashiwaya, S. & Ueda, M. Anomalous Josephson effect between even- and odd-frequency superconductors. *Phys. Rev. Lett.* **99**, 037005 (2007).
66. Tanaka, Y., Tanuma, Y. & Golubov, A. Odd-frequency pairing in normal-metal/superconductor junctions. *Phys. Rev. B* **76**, 054522 (2007).
67. Eldenstein, V. M. Triplet superconductivity and magnetoelectric effect near the *s*-wave-superconductor-normal-metal interface caused by local breaking of mirror symmetry. *Phys. Rev. B* **67**, 020505 (2003).
68. Li, X. S. *et al*. Large-area synthesis of high-quality and uniform graphene on copper foils. *Science* **324**, 1312-1314 (2009).
69. Novoselov, K. S. *et al*. Two-dimensional atomic crystals. *PNAS* **102**, 10451-10453 (2005).

**Supplementary Information** available at:
http://www.nature.com/article-assets/npg/ncomms/2017/170119/ncomms14024/extref/ncomms14024-s1.pdf

**Funding Acknowledgements**

The work was funded by the following agencies: Royal Society ('Superconducting Spintronics'), Leverhulme Trust (IN-2013-033), Schiff Foundation, the EPSRC (EP/N017242/1, EP/G037221/1, EP/K01711X/1, EP/K017144/1, EP/N010345/1, EP/M507799/1, EP/L016087/1), ERC Grant Hetero2D, EU Graphene Flagship, COST Action MP-1201, MSCA-IFEF-ST No. 656485-Spin3, Outstanding Academic Fellows programme at NTNU, Research Council of Norway (205591, 216700, and 24080).

**Author contributions**

J.W.A.R. supervised the project and with A.D.B conceived it. A.D.B. prepared the PCCO samples and performed structural and transport measurements with help from M.A. Single-layer graphene was grown and transferred onto PCCO by U.S., D.D.F. and M.B. STM measurements were performed by O.M., H. A. and Y. K. Raman analysis was carried out by A.K.O., D.Y., M.B., A.C.F. M.B. fabricated the Ag and Au nanoislands on single layer graphene on PCCO. The STM data were analysed by J.W.A.R., A.D.B. and O. M. with support from J.L. J.L. developed the theoretical model. The paper was written by A.D.B., O.M., A.C.F., J.L., J.W.A.R. All authors commented on the paper.

**Competing financial interest**
The authors declare no competing financial interests.